\documentclass[12pt,a4paper]{article}

\usepackage{amsmath}
\usepackage{amsfonts}
\usepackage{amsthm}
\usepackage[pdftex]{color,graphicx}
\usepackage{graphicx}
\usepackage{color}
\usepackage[colorlinks,linkcolor=blue,citecolor=red,urlcolor=black]{hyperref}

\usepackage{caption}
\usepackage{subcaption}

\newtheorem{Theorem}{Theorem}

\newtheorem{Remark}{Remark}
\theoremstyle{definition}

\newcommand{\dx}{{\mathrm{d}}x}
\newcommand{\dy}{{\mathrm{d}}y}

\newcommand{\dom}{{\mathrm{dom}~}}

\begin{document}

	\title{A note on twisted, straight, and sheared waveguide}

\author{Diana C. S. Bello \footnote{dianasuarez@estudante.ufscar.br} \\
{\small Departamento de Matemática – UFSCar, São Carlos, SP, 13560-970 Brazil}}

	\date{\today}
	
	\maketitle 
	
	\begin{abstract}
		
In this work, we analyze the Dirichlet Laplacian $-\Delta_{\Omega}^D$ in an unbounded waveguide $\Omega \subset \mathbb R^3$, 
where the cross-section is translated in a constant direction and rotated along a spatial line. We focus on the effects of twisting on the spectrum, discussing conditions under which discrete eigenvalues emerge. Our results highlight the interplay between geometry and spectral properties, showing that shearing can induce a richer spectral structure even in straight waveguides.

	\end{abstract}

	
	\
	
	
	
	%

	\section{Introduction}

The spectral analysis of the Laplacian operator in thin domains provides an effective framework for modelling electron dynamics in quantum waveguides, i.e.,
structures that constrain particle motion along specific directions, thereby creating a confined propagation channel. The geometry of the domain, together with the imposed confinement, plays a crucial role in shaping the system’s spectrum, influencing the existence and distribution of discrete eigenvalues of the operator.

In particular, let $\Omega$ be a waveguide in $\mathbb R^3$ and denote by $-\Delta_\Omega^D$ the Dirichlet Laplacian in $\Omega$. It is well known that if $\Omega$ is a bounded region, then this operator has a purely discrete spectrum. However, the situation changes when $\Omega$ is unbounded, and we can confirm that spectral properties can vary significantly depending on the geometry of $\Omega$. We explore this further in the next paragraph.

Consider the case where $\Omega$ is a twisted-bending waveguide. More precisely,
consider $\Gamma: \mathbb R \to \mathbb R^3$  as a $C^3$ spatial curve parameterized by its arc-length $x$,
and possessing an appropriate Frenet frame; denote by $k(x)$ and $\tau(x)$ its curvature and torsion at the point $\Gamma(x)$, respectively.
Let $S$ be a bounded open connected set in $\mathbb R^2$.
We define $\Omega$ as an unbounded waveguide obtained 
by moving the cross-section $S$ along the curve $\Gamma$ according to the Frenet referential.
At each point of $\Gamma$ the region
$S$ may also exhibit a (continuously differentiable) rotation angle $\alpha(x)$. 
Then, $\Omega$ is said to be bent if, and only if, $k \neq 0$; $\Omega$ 
is said to be twisted if, and only if, the cross section $S$ is non-rotationally invariant (i.e., $S$ is 
not a disc or an annulus centered at the origin of $\mathbb R^2$),
and $\tau - \alpha' \neq 0$. How these effects act on the spectrum of $-\Delta_\Omega^D$?
Without being exhaustive we recall some well known
results on this topic:

\begin{itemize}
	\item If $k  =\alpha=0$, i.e., $\Omega = \mathbb R \times S$ is a straight waveguide, the spectrum of $-\Delta_\Omega^D$ is purely essential; 
	$\sigma(-\Delta_\Omega^D) = \sigma_{ess} (-\Delta_\Omega^D) = [E, \infty)$, where $E$ is the first eigenvalue of the Dirichlet Laplacian in $S$.
	
	\item 
	If $\tau - \alpha'=0$, $k \neq0$ and $k(x) \to 0$, $|x| \to \infty$, one has
	$\sigma_{ess} (-\Delta_\Omega^D) = [E, \infty)$ and $\sigma_{dis}(-\Delta_\Omega^D)\neq \emptyset$; if one compares with the previous case, the assumptions do not
	change the essential spectrum but generate at least one discrete eigenvalue below $E$. On the other hand,
	if $k=0$, $\alpha \neq 0$ as compact support, and $S$ is a non-rotationally invariant cross section, it is possible to establish Hardy-type inequalities for $-\Delta_\Omega^D$,
	which avoid the existence of discrete spectrum. 
	Furthermore, if the bending is small compared to the twisting, there is no discrete spectrum.
	In this sense, one can say that bending acts as an attractive interaction and twisting acts as a repulsive one.
	For a survey on this topics, see \cite{brietnewref, briet, duclosfk, clark, duclos, pavelduclos, exnerseba, gold, davidtversusb, renger} 
	and references therein.
\end{itemize}

In this text, we explore the Dirichlet Laplacian operator in a ``twisted, straight, and sheared waveguide'', demonstrating that such a deformation can indeed generate discrete eigenvalues, revealing a richer spectral structure even among straight waveguides. In the following paragraphs, we present details of this model as well as our main result.

Denote by 
$\{e_1,e_2,e_3\}$ the canonical basis of  $\mathbb{R}^3$. Again, let $S \subset \mathbb{R}^2$ be a bounded open connected set in $\mathbb R^2$ so that it is
non-rotationally invariant.
Denote by $y :=(y_1,y_2)$ a point of $S$. Consider the function $f:\mathbb{R} \longrightarrow \mathbb{R}$, $f(x):= \beta x$, with $\beta \in [0, \infty)$.
Let $\alpha: \mathbb{R} \longrightarrow \mathbb{R}$ be a function of class $C^2$ with $\alpha' \in L^\infty(\mathbb{R})$. 
Now, consider $r_\beta: \mathbb{R} \longrightarrow \mathbb{R}^3$ the spatial curve
\begin{equation*}\label{defrefc}
		r_\beta(x):= (x, 0, \beta x), \quad x \in \mathbb{R}.
	\end{equation*}
Define the map 
\begin{equation}\label{lmasti}
	\begin{split}
		\mathcal{L}_{\beta}:\quad \mathbb{R} \times S \quad & \longrightarrow \mathbb{R}^3\\
		(x,y_1,y_2) & \longmapsto r_\beta(x)+(y_1 \cos(\alpha(x)) - y_2 \sin(\alpha (x))) e_2 + (y_1 \sin(\alpha (x)) + y_2 \cos(\alpha(x)))e_3,
	\end{split}
\end{equation}
and the region
\begin{equation*}\label{defot2}	
	\Omega_\beta := \mathcal{L}_\beta(\mathbb{R} \times S).
\end{equation*}

\begin{figure}[ht!]
	\centering
	\includegraphics[width=0.55\textwidth]{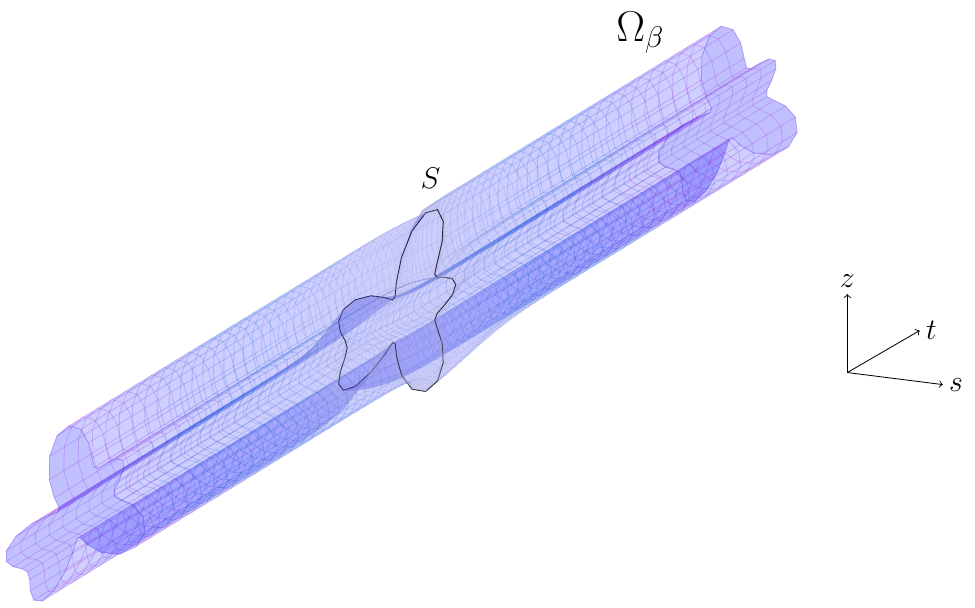}
	\caption{Twisted, straight, and sheared waveguide, with $\beta = 0.9$ and $\alpha(x) = (\tanh(x)+\pi)/2$.}
	\label{figrot1ti}
 \end{figure}

Geometrically, $\Omega_\beta$ is obtained by translating the region $S$ along the curve $r_\beta(x)$, such that, at each point of the curve, $S$ is parallel to the plane generated by ${e_2, e_3}$ and undergoes an angle rotation $\alpha(x)$ at the position $x$; see Figure \ref{figrot1ti}.  
In this context, the parameter $\beta$ represents a deformation that quantifies how much $\Omega_\beta$
deviates from the ``usual'' twisted, straight  waveguide $\Omega_0$.
Finally,
$\Omega_\beta$ will be referred to as a 
``twisted, straight, and sheared waveguide''.

Denote by $-\Delta^{D}_{\Omega_\beta}$ the Dirichlet Laplacian operator in $\Omega_\beta$, i.e., the self adjoint operator associated with 
the quadratic form   
	\begin{equation*} \label{fq311ti}
	\mathcal{Q}^{D}_{\Omega_\beta} (\psi) := \int_{\Omega_\beta} |\nabla \psi|^2 {\rm d}{\bf s}, \quad 
	\dom {\cal Q}^{D}_{\Omega_\beta} := \mathcal{H}^1_0 (\Omega_\beta),
\end{equation*}
where ${\bf s} = (s, t, z)$ denotes a point of $\Omega_\beta$, and $\nabla \psi$ the gradient of $\psi$.
	
We assume the condition
\begin{equation}\label{eq:assum1}
	\lim_{|x| \to \infty} \alpha^\prime (x) =:  0.
\end{equation}
This assumption implies that $\Omega_\beta$ is asymptotically straight at the infinity,
suggesting that the operator behaves in a predictable manner, thereby ensuring a certain stability of its essential spectrum.
This is confirmed by the result further below.

Denote $\partial_{y_1} := \partial/\partial y_1$, 	$\partial_{y_2} := \partial/\partial y_2$. 
Consider the auxiliary two-dimensional operator
\begin{equation}\label{eq:operadorst}
	T(\beta) := - \partial_{y_1}^2 - (1+ \beta^2) \partial_{y_2}^2,
\end{equation}
$\dom T(\beta)=\{v \in H_0^1(S):  T(\beta) v \in L^2(S)\}$;
denote by $E_1(\beta)$ its first eigenvalue. 
Since $T(\beta)$ is an elliptic operator with real coefficients, $E_1(\beta)$ is simple.

\begin{Theorem}\label{propressti}
	Assume the condition in \eqref{eq:assum1}. Then, for each $\beta \in [0, \infty)$, the essential spectrum of the operator
	$-\Delta^D_{\Omega_\beta}$ coincides with the interval $[E_1 (\beta),\infty)$. 
\end{Theorem}

The proof of Theorem \ref{propressti} is presented in Section \ref{essspec}.

The next step is to analyze
the discrete spectrum of $-\Delta_{\Omega_\beta}^D$. We start defining the constants 
\begin{equation*}\label{eq:constv}
    \begin{split}
  A_1 & :=\int_S y_2^2|\partial_{y_1} \chi|^2 \dy, \quad A_2 :=\int_S y_2|\partial_{y_1} \chi|^2 \dy, \quad A_3 :=\int_S |\partial_{y_1} \chi|^2 \dy,   \\
  B_1 &:=\int_S y_1^2|\partial_{y_2} \chi|^2 \dy, \quad B_2 :=\int_S y_1|\partial_{y_2} \chi|^2 \dy, \quad B_3 :=\int_S |\partial_{y_2} \chi|^2 \dy,\\
  C_1 &:=\int_S y_1 y_2 \partial_{y_1} \chi \partial_{y_2} \chi \dy, \quad C_2 :=\int_S  y_2 \partial_{y_1} \chi \partial_{y_2} \chi \dy, \quad C_3 :=\int_S  y_1 \partial_{y_1} \chi \partial_{y_2} \chi  \dy, \quad C_4 :=\int_S  \partial_{y_1} \chi \partial_{y_2} \chi \dy,  
\end{split}
\end{equation*}
and the function
$V: \mathbb{R} \longrightarrow \mathbb{R}$,
\begin{align*}
	V(x) & :=(A_1 + B_1 - 2 C_1) (\alpha^\prime (x))^2 +2(C_3 -A_2)\beta \alpha^\prime (x) \sin(\alpha (x)) + 2(B_2-C_2) \beta \alpha^\prime (x) \cos(\alpha (x))\\
& + (A_3-B_3) \beta^2 (\sin(\alpha (x)))^2 + 2 C_4 \beta^2 \sin(\alpha (x))\cos(\alpha (x)).
\end{align*}

If $\beta =0$, it is well know that $-\Delta_{\Omega_0}^D$ has no discrete eigenvalues; see for example \cite{brietnewref}.
Then, we present our main contribution on the theme.
\begin{Theorem}\label{exidisspcti}
	Take $\beta \in (0, \infty)$. Suppose that there exists a non null continuous function  $\alpha: \mathbb R \to \mathbb R$ satisfying the condition in
	\eqref{eq:assum1}, so that  $V(x) \in L^1(\mathbb{R})$ and $\int_\mathbb{R} V(x) \dx <0$. Then, $\inf \sigma (-\Delta_{\Omega_\beta}^D) < E_1(\beta)$, i.e., 
	$\sigma_{dis}(-\Delta_{\Omega_\beta}^D) \neq \emptyset$. 
\end{Theorem}

The proof of Theorem \ref{exidisspcti} is presented in Section \ref{disspec}. 
In Section \ref{resnum} we present a particular example by considering the cross section $S=(0,\pi)^2$. The goal is to show that the number of discrete eigenvalues can vary significantly depending on the function $\alpha(x)$ and the parameter $\beta$.

\begin{Remark}\label{remidn}{\rm 
The model built in this work was inspired by \cite{ davidk, verri}. In these papers, the shearing effect was introduced for the first time.}
\end{Remark}

\begin{Remark}\label{remidn2}{\rm 
		 Look at the definition of $V(x)$. Remark that 
		 $A_1 + B_1 - 2 C_1 = \int_S \left(y_2 \partial_{y_1} \chi -  y_1 \partial_{y_2} \chi \right)^2 \dy  \geq 0$. Then,
		 the negative value of the integral $\int_\mathbb{R} V(x) \dx$ depends explicitly of the parameter $\beta$; 
		 in case $\beta=0$, one has $\int_\mathbb{R} V(x) \dx \geq 0$.}
\end{Remark}

	This work is organized as follow. In Section \ref{essspec} we study  the essential spectrum of  $-\Delta_{\Omega_\beta}^D$. Section \ref{disspec} is dedicated to study the discrete spectrum of the operator. In Section \ref{resnum} is presented a simple numerical analysis and some considerations.
	
\section{Essential spectrum}\label{essspec}

In this section, we study the essential spectrum for $-\Delta^D_{\Omega_\beta}$.
For each $\beta \in (0, \infty)$, define the spatial curve
$\tilde{r}_\beta(x): = (x, 0, \beta x)$, $x \in \mathbb{R}$, and the mapping
$\tilde{{\cal L}}_\beta(x, y_1, y_2):=  \tilde{r}_\beta(x) + y_1 e_2 + y_2 e_3$, $(x, y_1, y_2) \in \mathbb{R} \times S$. Consider the straight waveguide
$\tilde{\Omega}_\beta := \tilde{\mathcal{L}}_\beta(\mathbb{R} \times S)$.
Proposition 3 of \cite{verri} shows that the Dirichlet Laplacian in $\tilde{\Omega}_\beta$ has a purely essential spectrum and it is equal to the interval
$[E_1(\beta), \infty)$. Since the essential spectrum of the Dirichlet Laplacian in tubular domains is determined 
by the geometry of the region at infinity only, the statement of Theorem \ref{propressti} in the Introduction is expected. 
However, we  describe the mains ideas of its proof below.

\vspace{0.3cm}

\noindent
{\bf Proof of Theorem \ref {propressti}:}
Let $K \subset \Omega_\beta$ be a compact set so that $\Omega_{\mathrm{ext}} := \Omega_\beta \backslash K = \Omega_{\mathrm{ext}}^1 \cup \Omega_{\mathrm{ext}}^2$ (disjoint union), where
$\Omega_{\mathrm{ext}}^1$ and  $\Omega_{\mathrm{ext}}^2$ are isometrically affine to a straight half-waveguide. Define $\Omega_{\mathrm{inter}} := int(K)$.
Consider the quadratic form 
$Q_{\mathrm{inter}}^{DN} \oplus Q_{\mathrm{ext}}^{DN}$, where
\[Q_{\omega}^{DN} (\psi) = \int_{\Omega_{\omega}}  |\nabla \psi|^2 {\rm d}{\bf x},\]
\[\mathrm{dom}~ Q^{DN}_{\omega} = \{\psi \in H^1 (\Omega_{\omega}): \psi = 0 \,\, {\rm in} \,\, \partial \Omega_ \beta \cap \partial \Omega_{\omega} \},\]
$\omega \in \{\mathrm{inter}, \mathrm{ext}\}$. 
Denote by  $H_{\mathrm{inter}}^{DN}$ and $H_{\mathrm{ext}}^{DN}$ the self-adjoint operators associated with $Q_{\mathrm{inter}}^{DN}$ and $Q_{\mathrm{ext}}^{DN}$, respectively. 
One has  
\begin{equation*}
	-\Delta_{\Omega_\beta}^D  \geq H_{\mathrm{inter}}^{DN} \oplus H_{\mathrm{ext}}^{DN},
\end{equation*}
in the quadratic form sense. 

Proposition 3 and Remark 1 in \cite{verri} imply that $\sigma_{ess}(H_{\mathrm{ext}}^{DN})= [E_1(\beta), \infty)$. By minimax principle, and since the spectrum of 
$H_{\mathrm{inter}}^{DN}$ is purely discrete, we have
\begin{equation*}
	\inf \sigma_{ess} (-\Delta_{\Omega_\beta}^D)  \geq \inf \sigma_{ess}(H_{\mathrm{inter}~}^{DN} \oplus H_{\mathrm{ext}}^{DN}) =  \inf \sigma_{ess}(H_{\mathrm{ext}}^{DN}) = E_1(\beta),
\end{equation*}
i.e., $\sigma_{ess} (-\Delta_{\Omega_\beta}^D) \subseteq [E_1(\beta), \infty)$.

Finally, by the considerations in \cite{verri}, it is possible to construct a Weyl sequence supported in $\Omega_{\mathrm{ext}}$ associated with each  $\lambda \geq E_1(\beta)$.
Then, one has  $[E_1(\beta), \infty) \subseteq \sigma_{ess} (-\Delta_{\Omega_\beta}^D)$.  \qed

	\section{Discrete spectrum}\label{disspec}

   In this section, we study the existence of the discrete spectrum for $-\Delta_{\Omega_\beta}^D$. First, we fix some notation that will be useful in what follows. Let $Q$ be a closed, lower-bounded sesquilinear form with domain $\mathrm{dom}~ Q$ dense in a Hilbert space $H$. Let $A$ be the self-adjoint operator associated with $Q$. The Rayleigh quotient of $A$ can be defined as
   \begin{equation}\label{rayquoin}
   	\lambda_j(A) = \inf_{\substack{G \subset \mathrm{dom}~ Q \\ \dim G = j}}   
   	\sup_{\substack{ \psi \in G \\ \psi \neq 0}}  
   	\frac{Q(\psi)}{\|\psi\|^2_H}.
   \end{equation}
   Let $\mu = \inf \sigma_{\mathrm{ess}}(A)$. The sequence $\{\lambda_j(A)\}_{j \in \mathbb{N}}$ is non-decreasing and satisfies:
   \begin{itemize}
   	\item[(i)] If $\lambda_j(A) < \mu$, then this is a discrete eigenvalue of $A$;
   	\item[(ii)] If $\lambda_j(A) \geq \mu$, then $\lambda_j(A) = \mu$;
   	\item[(iii)] The $j$-th eigenvalue of $A$ below $\mu$ (if it exists) coincides with $\lambda_j(A)$.
   \end{itemize}

\subsection{Change of coordinates}

In this section we perform a change of coordinates to work in the domain $\Lambda := \mathbb R \times S$.

Recall the map  ${\cal L}_\beta$ given by \eqref{lmasti} in the Introduction so that $\Omega_\beta = \mathcal{L}_\beta(\Lambda)$.
By Proposition  1  in \cite{verri}, ${\cal L}_\beta$ is a local diffeomorphism $C^{0,1}$.
Since  ${\cal L}_\beta$ is injective, we obtain a global diffeomorphism  $C^{0,1}$. Then, the region
$\Omega_\beta$ can be identified to the Riemannian manifold $(\Lambda, G_\beta)$, where $G_\beta=(G_\beta^{ij})$ is 
metric induced by ${\cal L}_\beta$, i.e.,
\begin{equation*}
	G_\beta^{ij} = \langle {\cal G}_\beta^i, {\cal G}_\beta^j \rangle = G_\beta^{ji}, \quad i,j=1,2,
\end{equation*}
where
\begin{equation*}
	{\cal G}_\beta^1 = \frac{\partial {\cal L}_\beta}{\partial x}, \quad 
{\cal G}_\beta^2 = \frac{\partial {\cal L}_\beta}{\partial y_1}, \quad {\cal G}_\beta^3 = \frac{\partial {\cal L}_\beta}{\partial y_2} .
\end{equation*}

More precisely, we define
\begin{equation*}
    M(x,y) := K(x,y)\cos(\alpha(x)) + P(x,y)\sin(\alpha(x)), \quad \mbox{and} \quad
    N(x,y) := -K(x,y)\sin(\alpha(x)) + P(x,y)\cos(\alpha(x)),
\end{equation*}
where
\begin{equation*}
        K(x,y) := -\alpha^\prime(x) y_1 \sin(\alpha (x)) - \alpha^\prime(x) y_2 \cos(\alpha (x)), \quad
    L(x,y) := \alpha^\prime(x) y_1 \cos(\alpha (x)) - \alpha^\prime(x) y_2 \sin(\alpha (x)) +\beta.\\
\end{equation*}
Then,
\begin{equation*}
G_\beta  = \nabla {\cal L}_\beta \cdot (\nabla {\cal L}_\beta)^t = \left(
\begin{array}{ccc}
	1 +  K^2(x,y) + P^2(x,y) &  M(x,y)   &  N(x,y)  \\
	M(x,y)                              & 1       &  0 \\
	N(x,y)                                 &    0                & 1
\end{array} \right) \quad \mbox{and} \quad
 \det G_\beta = 1.
\end{equation*}

Now, we consider the unitary operator
\begin{equation*}\label{unituvar}
	\begin{array}{llll}
		{\cal U}_\beta: &   L^2(\Omega_\beta)  &  \to &  L^2(\Lambda) \\
		&    \psi   &  \mapsto  &        \psi \circ {\cal L}_\beta
	\end{array}.
\end{equation*}

Finally, we define
\begin{align*}\label{compaquadravarti}
    Q_\beta(\psi) & :=  Q_{\Omega_\beta}^{D}({\cal U}_\beta^{-1} \psi)  
	= \int_{\Lambda} \langle \nabla \psi, G_\beta^{-1} \nabla \psi \rangle \sqrt{{\rm det}\, G_\beta} \, {\rm d}x {\rm d}y  \nonumber \\ 
	& =  \int_{\Lambda} \bigg( \left| \psi^\prime + (\alpha^\prime (x) y_2 - \beta \sin(\alpha (x))) \frac{\partial \psi}{\partial y_1} - (\alpha^\prime (x) y_1 + \beta \cos(\alpha (x))) \frac{\partial \psi}{\partial y_2} \right|^2 +  |\nabla_y \psi|^2 \bigg) {\rm d}x {\rm d}y,
\end{align*}
$\dom Q_\beta = {\cal U}_\beta ({\cal H}_0^1(\Omega_\beta))$, where
$\psi' := \partial \psi / \partial x$,  and $\nabla_y \psi := (\partial_{y_1} \psi, \partial_{y_2} \psi)$. 
Denote by $H_\beta$ the self-adjoint operator associated with $Q_\beta(\psi)$.

\subsection{Proof of the results}
	
Now, we have conditions to prove Theorem \ref{exidisspcti}. 
	
\vspace{0.2cm}
\noindent
{\bf Proof of Theorem \ref{exidisspcti}:}
Consider the quadratic form
\begin{equation*}
q_\beta (\psi) := Q_\beta(\psi) - E_1(\beta)\|\psi\|_{L^2(\Lambda)}^{2}, \quad \dom q_\beta = \dom Q_\beta.
\end{equation*}
By \eqref{rayquoin} and Theorem \ref{propressti}, it is enough to find out a non null function $\psi \in \dom q_\beta$ so that $q_\beta(\psi) < 0$. 
For that, take $w \in C^\infty(\mathbb{R})$ a function so that $w(x)=1$, for all  $x \in [-1,1]$, and 
 $w(x)=0$, for all $x \in \mathbb{R} \setminus (-2,2)$. Then, for each  $n \in \mathbb{N}\setminus \{0\}$, define
\begin{equation*}
w_n(x) := w \left(\frac{x}{n}\right) \quad \mbox{and} \quad \psi_n(x,y) := w_n(x) \chi (y),
\end{equation*}  
where $\chi$ denotes the normalized eigenfunction correspondingly to the first eigenvalue $E_1(\beta)$ of the auxiliary operator $T(\beta)$. 
In particular,
\begin{equation}\label{estdericuoffti}
		\int_{\mathbb{R}} |w'_n|^2 \dx = \frac{1}{n} \int_{\mathbb{R}} |w'|^2 \dx \to 0, \quad \hbox{as} \quad n \to \infty,
\end{equation}
and
\begin{equation}\label{firsteigcrossti}
E_1(\beta)  =	\int_S \bigg(|\partial_{y_1} \chi|^2 +  (1 + \beta^2)  |\partial_{y_2} \chi|^2 \bigg) \dy.
\end{equation}

By \eqref{firsteigcrossti}, and since $\int_S \chi \partial_{y_2} \chi \dy = \int_S  \chi \partial_{y_1} \chi \dy =0$, and $\int_S y_1\chi \partial_{y_2} \chi \dy = \int_S  y_2\chi \partial_{y_1} \chi \dy =0$,  one has
\begin{equation*}
\begin{split}
q_\beta(\psi_n) & = Q_\beta(\psi_n) - E_1(\beta)\|\psi_n\|_{L^2(\Lambda)}^{2}\\
& = \int_{\Lambda} \bigg( \left| w_n^\prime \chi + (\alpha^\prime (x) y_2 - \beta \sin(\alpha (x))) w_n \partial_{y_1} \chi - (\alpha^\prime (x) y_1 + \beta \cos(\alpha (x))) w_n \partial_{y_2} \chi \right|^2\\
&  + \left|w_n \right|^2 \left|\nabla_{y} \chi \right|^2 \bigg)   \dx\dy - E_1(\beta) \int_{\Lambda} \left|w_n \right|^2 \left|\chi \right|^2 \dx\dy\\
& = \int_{\Lambda}\bigg[ |w^\prime_n|^2 |\chi|^2   + 2 (\alpha^\prime (x) y_2 - \beta \sin(\alpha (x)))  w_n w_n^\prime \chi \partial_{y_1} \chi -  2 (\alpha^\prime (x) y_1 - \beta \cos(\alpha (x)))  w_n w_n^\prime \chi \partial_{y_2} \chi\\
&  + |w_n|^2 
\bigg([1+ (\alpha^\prime (x) y_2 - \beta \sin(\alpha (x)))^2]|\partial_{y_1} \chi|^2   + [1+(\alpha^\prime (x) y_1 + \beta \cos(\alpha (x)))^2] |\partial_{y_2} \chi|^2 \\
& - 2 (\alpha^\prime (x) y_2 - \beta \sin(\alpha (x)))(\alpha^\prime (x) y_1 + \beta \cos(\alpha (x))) \partial_{y_1} \chi \partial_{y_2} \chi - E_1(\beta) \chi|^2 \bigg) \bigg] \dx\dy\\
& = \int_{\Lambda}\bigg[ |w^\prime_n|^2 |\chi|^2    + |w_n|^2 
\bigg( (\alpha^\prime (x) y_2 - \beta \sin(\alpha (x)))^2|\partial_{y_1} \chi|^2   + [(\alpha^\prime (x) y_1 + \beta \cos(\alpha (x)))^2 - \beta^2] |\partial_{y_2} \chi|^2 \\
& - 2 (\alpha^\prime (x) y_2 - \beta \sin(\alpha (x)))(\alpha^\prime (x) y_1 + \beta \cos(\alpha (x))) \partial_{y_1} \chi \partial_{y_2} \chi \\
& 
+ |\partial_{y_1} \chi|^2 + (1+\beta^2) |\partial_{y_2} \chi|^2- E_1(\beta) |\chi|^2 \bigg) \bigg] \dx\dy\\
& = \int_\mathbb{R} |w'_n|^2  \dx + \int_\mathbb{R} V(x) |w_n|^2\dx.
\end{split}
\end{equation*}
Finally, by \eqref{estdericuoffti}, and since $\|w_n\|_\infty \leq 1$, and  $\int_\mathbb{R} V(x)dx <0$, it follows that 
\begin{equation*}
q_\beta(\psi_n) \to \int_\mathbb{R} V(x)dx, \quad \mbox{as} \quad n \to \infty.	
\end{equation*}		
Now, just to take  $N \in \mathbb{N}$ large enough so that $q_\beta(\psi_N) <0$.  \qed
\vspace{0.3cm}


\section{A numerical example}\label{resnum}

\begin{figure}[ht!]
	\centering
	\includegraphics[width=0.55\textwidth]{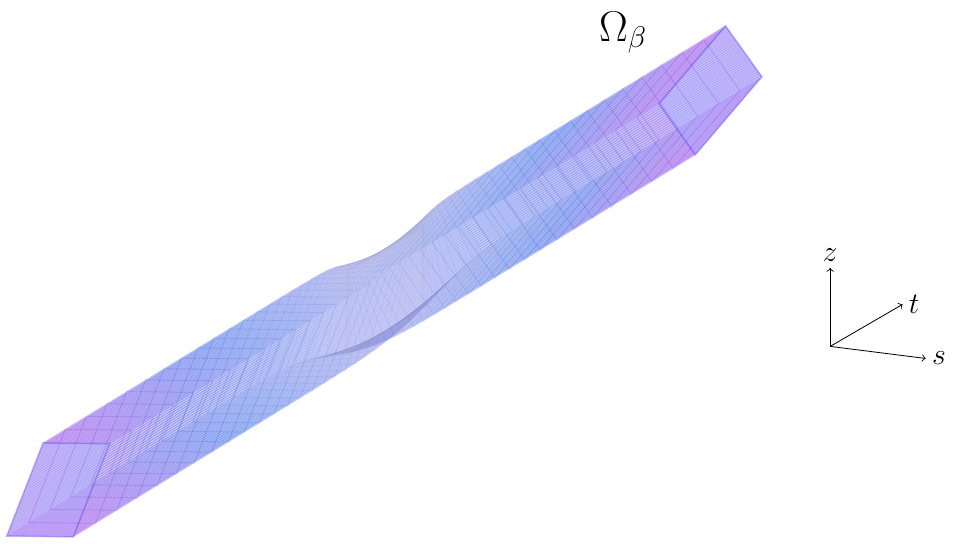}
	\caption{Twisted, straight and sheared waveguide, with $\beta=0.9$ and $\alpha(x)=(\tanh(x)+\pi)/2$.}
	\label{figroti_sq}
\end{figure}

In this section, we consider the particular case where the cross section $S$ is the square $(0,\pi)^2$, and 
\[\alpha(x) = c \tanh(x) + \pi/2,\]
for some constant $c \in \mathbb R$.
Note that $\alpha(x)$ is a $C^\infty$ function  and  $\alpha^\prime (x)=c(1-\tanh^2(x))$ satisfies the condition \eqref{eq:assum1}. 
The proposal is to analyze different values of $c$ and $\beta$ to show that the results can change significantly according to them.

Recall the auxiliary operator $T(\beta)$ given by \eqref{eq:operadorst}. Since $S=(0, \pi)^2$, the
first eigenvalue of $T(\beta)$  is $E_1(\beta)=2+\beta^2$, and the correspondingly eigenfunction is $\chi(y) = 2\sin(y_1) \sin(y_2)/\pi$. Furthermore,
\begin{align*}
A_1 &  = \int_S y_2^2\left|\frac{2}{\pi}\cos(y_1) \sin(y_2) \right|^2 \dy = B_1 =\int_S y_1^2\left|\frac{2}{\pi}\sin(y_1) \cos(y_2) \right|^2 \dy = \frac{1}{6}(2\pi^2 -3) \approx 2.7899,\\
A_2 &  = \int_S y_2 \left|\frac{2}{\pi}\cos(y_1) \sin(y_2) \right|^2 \dy = B_2 =\int_S y_1 \left|\frac{2}{\pi}\sin(y_1) \cos(y_2) \right|^2 \dy = \frac{\pi}{2},\\
A_3 &  = \int_S  \left|\frac{2}{\pi}\cos(y_1) \sin(y_2) \right|^2 \dy = B_3 =\int_S  \left|\frac{2}{\pi}\sin(y_1) \cos(y_2) \right|^2 \dy = 1,\\
C_1 &= \int_S  \frac{4}{\pi^2} y_1 y_ 2 \cos(y_1) \sin(y_2) \sin(y_1) \cos(y_2)  \dy = \frac{1}{4},\\
C_2 & = C_3 = C_4 = 0, 
\end{align*}
and 
\begin{equation*}
	V(x) = \left( \frac{2}{3} \pi^2 - \frac{3}{2}\right)   (\alpha^\prime (x))^2 + \pi \beta \alpha^\prime (x) \left(\cos(\alpha (x)) - \sin(\alpha (x)) \right), \quad \forall x \in \mathbb{R}.	
\end{equation*}

In the next, all the numerical analyses where obtained from the {\it Freefem++} software \cite{freefem}.
The number of discrete eigenvalues of $-\Delta_{\Omega_\beta}^D$ will be denoted by 
$\# {\cal N} (\beta)$.

\begin{figure}[ht!]
	\centering
	\begin{subfigure}[b]{0.48\textwidth}
		\centering
	\includegraphics[width=0.9\textwidth]{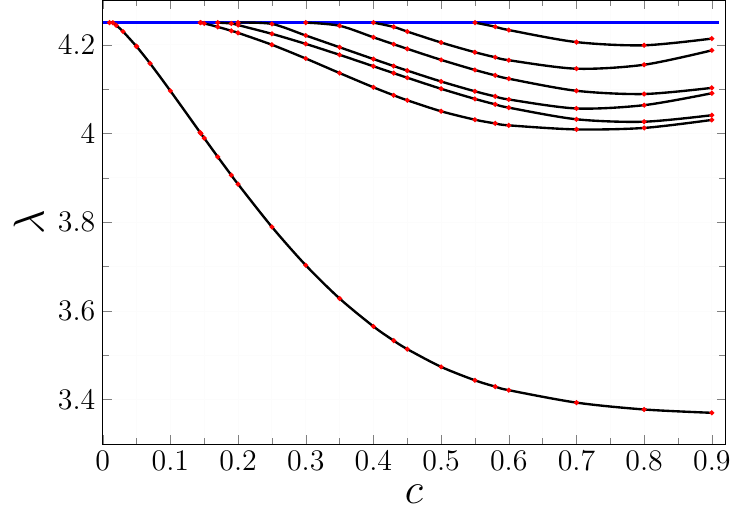}
	\caption{$\lambda$ vs constant $c$.}
	\label{fig:eigenv}
	\end{subfigure}
	\hfill
	\begin{subfigure}[b]{0.51\textwidth}
		\centering
	\includegraphics[width=0.9\textwidth]{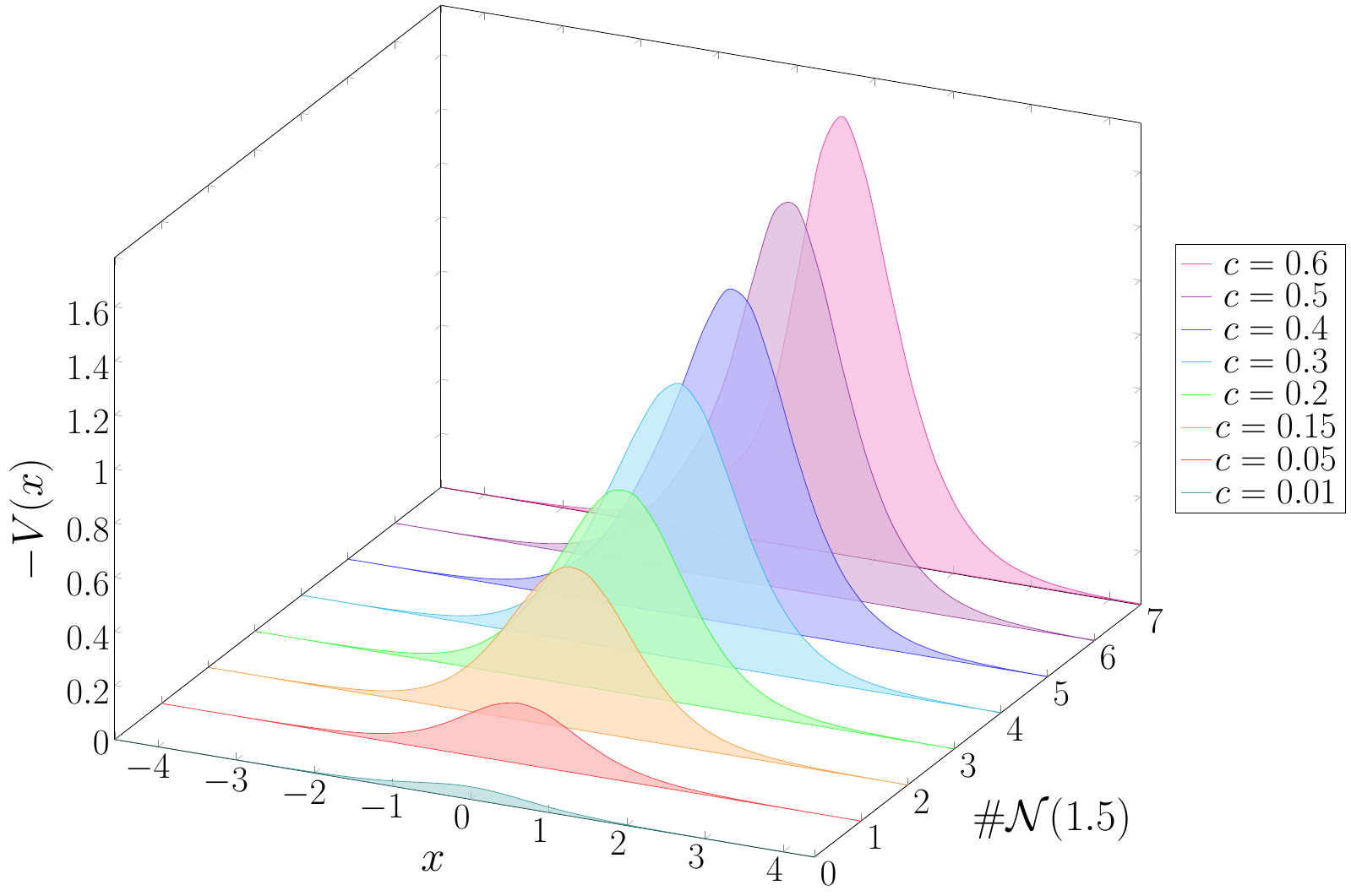}
	\caption{ $-V(x)$ vs $\# {\cal N}(1.5)$.}
	\label{fig:cascade}
	\end{subfigure}
	\caption{Discrete eigenvalues of $-\Delta_{\Omega_{1.5}}^D$.}
	\label{fig:two}
\end{figure}

\vspace{0.2cm}
\noindent
{\bf Case I.} ($\beta$ fixed)
Fix  $\beta = 1.5$. Then, $E_1(1.5) = 4.25$ is the infimum of the essential spectrum of $-\Delta_{\Omega_{1.5}}^D$ (see blue line in the  Figure \ref{fig:eigenv}).
We analyze the relation between $\int_{\mathbb{R}}V(x) \dx$ and  $\# {\cal N}(1.5)$; note that $V(x)$ depends on $c$.
From a numerical analysis, we obtained the following considerations:

\vspace{0.2cm}

- For each $c \in (0.01,0.144]$, one has $\# {\cal N}(1.5)=1$; see Figure \ref{fig:eigenv}.

\vspace{0.2cm}

- If $c>0.144$, then $\# {\cal N}(1.5)$ increases when $|\int_\mathbb{R} V(x) \dx|$ increases; see Figure \ref{fig:cascade}.

\vspace{0.2cm} 

\begin{figure}[ht!]
\centering
\includegraphics[width=0.5\textwidth]{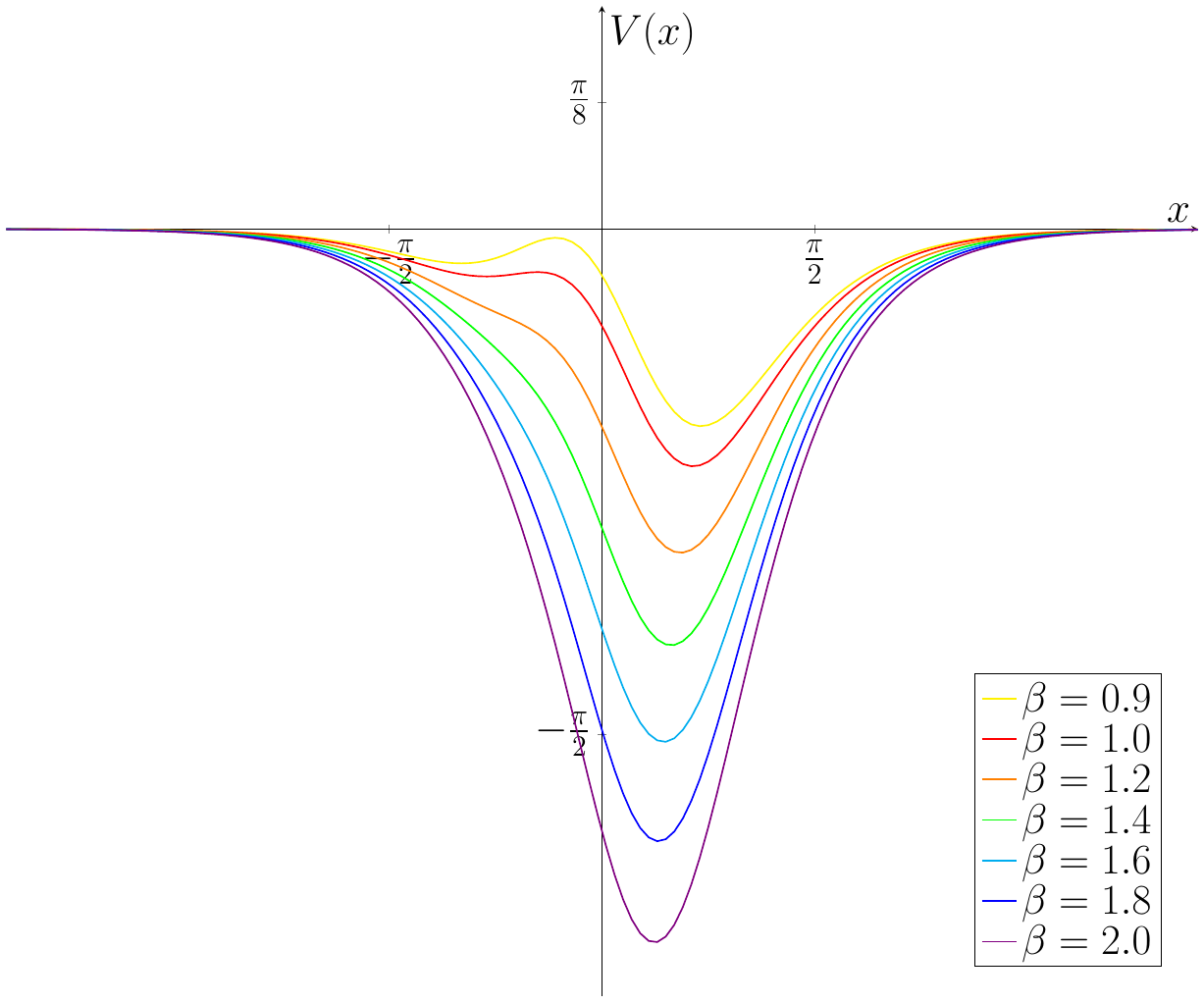}
\caption{Graph of $V(x)$ with $c=0.5$, and some values of $\beta$.}
\label{functionVx}
\end{figure}

\vspace{0.2cm}
\noindent
{\bf Case II.} ($c$ fixed)
Fix $c=0.5$. Then, $\alpha(x) = (\tanh(x)+\pi)/2$ and  $\Omega_\beta$ is represented in Figure \ref{figroti_sq}.  
As we can see by Figure \ref{functionVx}, for different values of $\beta$ one has $V(x)<0$, for all $x \in \mathbb{R}$, i.e., 
$\int_\mathbb{R} V(x)<0$. Then, $\# {\cal N}(\beta) > 0$.
Furthermore,  Figures \ref{fig:cascade} and \ref{functionVx} show that the value of $\beta$ can 
affect the number of discrete eigenvalues.

	\vspace{0.2cm}
	\noindent
	{\bf Acknowledgments}

	\vspace{0.2cm}
	\noindent

The author is grateful to Alessandra Verri for useful suggestions, discussions, very detailed and critical reading of the manuscript. The work has been supported by CAPES (Brazil) through the process: 88887.511866/2020-00.

\vspace{0.2cm}
	\noindent
	
	\bibliographystyle{plain}

	\end{document}